
\magnification=1200
\hsize=17 true cm
\vsize=22 true cm
\baselineskip 15 pt

\def\llam{$\lambda\lambda$}
\def\lam{$\lambda$}
\def\a{${\rm \AA}$}

\def\topmongofig#1#2{%
\topinsert
\includegraphics{#1}
\vskip 4.8 truein
$$\vbox{\hsize=5.5truein%
\noindent{\bf Figure #2. }
}$$
\endinsert
}

\def\pagemongofig#1#2{%
\pageinsert
\includegraphics{#1}
\vskip 6.8 truein
$$\vbox{\hsize=5.5truein%
\noindent{\bf Figure #2. }
}$$
\endinsert
}

\def\tabmongofig#1{%
\pageinsert
\includegraphics{#1}
\vskip 6.8 truein
$$\vbox{\hsize=5.5truein
}$$
\endinsert
}

\def\ref {\par \noindent \parshape=6 0cm 12.5cm
 0.5cm 12.5cm 0.5cm 12.5cm 0.5cm 12.5cm 0.5cm 12.5cm 0.5cm 12.5cm}
\vtop to 2.0 true cm {}
\centerline {\bf DETERMINATION OF ATMOSPHERIC PARAMETERS OF T TAURI STARS}
\vskip 1.0 true cm
\centerline { R. Piorno Schiavon$^{1,2}$,  C. Batalha$^1$, B. Barbuy$^2$}
\bigskip
\bigskip
\centerline {1 Observat\'orio Nacional, Departamento de Astrof\'\i sica}

\centerline {Rua General Jos\'e Cristino, 77, S\~ao Crist\'ov\~ao}

\centerline {20921-400 Rio de Janeiro, Brazil}
\bigskip
\centerline {2 Universidade de S\~ao Paulo, IAG, Departamento de Astronomia}

\centerline { C.P. 9638, S\~ao Paulo 01065-970, Brazil}
\vskip 3.0 true cm
\noindent January, 1995.

\noindent To appear in Astronomy and Astrophysics (Main Journal)

\noindent Send requests to: R. Piorno Schiavon (address 2)


\vfill\eject

\noindent {\bf ABSTRACT.} The inferred effective temperatures (T$_{\rm
eff}$) and surface gravities of T Tauri stars (TTS) are usually
contaminated by the presence of a non stellar continuum emission
(veiling) and the strong chromospheric activity characteristic of these
objects. In this work, we develop a method to determine T$_{\rm eff}$'s
and surface gravities (log$\,g$) of a group of TTS.  This method is
based on the comparison between observed and theoretical molecular and
atomic line depth ratios being therefore insensitive to the influence
of veiling. We show the strong dependence of our line depth ratios upon
gravity and temperature.  The resulting gravities, as expected for TTS,
average between the values of dwarf and giant stars. Previously
published gravities for each of our stars vary widely due in part to
the differences in the adopted visual extinction, veiling (if ever
considered) and methods of assessing the stellar luminosity. Our values
of T$_{\rm eff}$ and log$\,g$ have the uniqueness of being entirely
derived from high resolution data and are not affected by circumstellar
extinction or veiling when line ratios are used. They therefore serve
as more reliable input parameters for future spectral synthesis
analyses of T Tauri Stars requiring model atmospheres. We provide a
table relating theoretical line depth ratio with T$_{\rm eff}$ and log
$\,g$ for easy assessment of TTS fundamental parameters.

\bigskip
\noindent {\bf 1. INTRODUCTION}
\vskip 0.5 true cm
Reliable atmospheric parameters of T Tauri stars
(TTS) provide the cornerstone in understanding the formation and
pre main sequence evolution of low mass stars as they contract towards
the zero age main sequence. Without the basic information of effective
temperatures (T$_{\rm eff}$) and surface gravities (log$\,g$), the
accurate determination of the elemental atmospheric abundances of TTS
is likely to result in failure.

The present picture of a classical T Tauri star (CTTS) consists of a
central star surrounded by an optically thick accreting disk rotating
at about the stellar equatorial plane. First attempts to model this
system were made by Hartmann \& Kenyon (1987) and Bertout, Basri \&
Bouvier (1988), in which a shear boundary layer interfacing the stellar
photosphere and the disk produces the continuum veiling and part of the
broad lines typical of these stars. In more recent approaches, the
magnetosphere has an active role in shaping the circumstellar
environment. Depending on the strength of the stellar magnetic field
lines, the disk may never reach the star, being disrupted at about the
corotational radius (K\"onigl 1991, Shu et al. 1994, Kenyon et al.
1994). In this case, the disk matter is funneled along the magnetic
field lines to the stellar surface, generating hot gas at the
footpoints of the magnetic columns.

Owing to the above mentioned phenomena, the spectrum of a CTTS is
usually very complex and its main features are: {\it i)} infrared (IR)
and radio flux excesses, which are largely believed to be disk
emission; {\it ii)} optical and ultraviolet excesses which correlate
with the IR fluxes  (Bertout 1989, Hartigan et al. 1990). This
nonstellar continuum radiation in the optical range reduces the depth
of photospheric absorption lines and is the main source of the so
called veiling (Basri \& Batalha 1990, Hartigan et al. 1991); {\it
iii)} chromospheric filling in of selective absorption lines, which
constitutes a small fraction of the veiling. This property is well
correlated with stellar activity (Finkenzeller \& Basri 1987). Further
indications of stellar activity are the {\it iv)} X-ray emission,
probably due to hot solar-type coronae and {\it v)} the periodical
photometric variability suggesting the transit of bright and dark spots
on the stellar atmosphere (Bouvier 1990). {\it vi)} The upper Balmer
series are in emission indicating  mass transport in the stellar
surroundings.  {\it vii)} Forbidden lines are present and their total
flux correlates with the IR luminosity  (Cabrit et al. 1990).

The so called weak T Tauri stars (WTTS) differ from the CTTS in their
observed characteristics, presenting  much milder near IR and blue
color excesses (if any), no forbidden line emission (Edwards et al.
1994), the absence of a Balmer jump as well as Balmer line emission
(Valenti, Basri \& Johns 1993). The current interpretation of these
features is the lack of accreting disks around the  WTTS.

The application of canonical methods to determine the atmospheric
parameters of CTTS is largely prohibited by the above spectral
features, although they might be effective if applied to WTTS. First of
all, color indices are severely changed by the \lam -dependent
nonstellar emission and circumstellar extinction, misleading any of the
classical attempts to infer T$_{\rm eff}$ or metallicities from the
usual photometric indicators. Moreover, the optical veiling and
probably the chromospheric filling in reshape the wings of moderate to
strong lines inhibiting the use of the synthetic profile technique to
derive the stellar T$_{\rm eff}$. The same argument can be presented
against using the classical curve of growth method   to analyze
the  properties of CTTS. The determination of TTS surface gravities
from their locus on the theoretical HR-diagram is strongly affected by
the errors in luminosity, temperature and the uncertainties on the
pre-main-sequence evolutionary tracks.

Recently, however, this situation has begun to change as more analyses
based on high resolution data have been presented.  High resolution and
high S/N spectroscopy has allowed direct measurements of the intrinsic
stellar luminosity, after  proper veiling correction are carried out
(Hartigan et al. 1989, Basri \& Batalha 1990 and Hartigan et al.
1991). This is done by comparing the TTS spectrum with that of a
template star and analyzing the residuals. Implicitly, it is assumed
that both stars share similar atmospheric parameters, metallicity and
chromospheric activity.

Theoretical analyses of the underlying photospheres have been made by
Calvet, Basri \& Kuhi (1984) and Calvet \& Marin (1987). In these
works, the absorption spectra of classical TTS are synthesized using  a
simplified NLTE treatment. The temperature minimum and the associated
gas column density are the key parameters in matching the
observations.  The curve of growth method has also been successfully
applied to Sz 19 and its fundamental stellar  parameters have been
derived (Franchini et al. 1991). In these studies,  the veiling as an
external source of continuum emission has not been considered because
of the intrinsic theoretical approach and because the star lacks
veiling. Specific studies of atmospheric features where consistent
veiling corrections are carried out are also being released. Basri,
Martin \& Bertout (1991) derive lithium abundances for a large sample
of TTS and find abnormally large  values of N$_{Li}$ ($>$ 3.3) for
some of the stars.  They suggest that the overabundance might be the
result of errors in the ascribed stellar parameters. Magazz\`u, Rebolo
\& Pavlenko (1992) found that Li abundances are reduced, on the
average, when NLTE effects are taken into account. Nevertheless, some
of their stars still remain overabundant in Li. Martin et al.  (1994)
performed NLTE analyses of more than 50 WTTS and found no Li
overabundances. They conclude that the overabundances found in CTTS may
be attributted to the errors in the input atmospheric parameters.
Thus, an accurate method to determine CTTS atmospheric parameters can
help to resolve the Li problem in TTS.

Batalha \& Basri (1993) and Batalha et al.  (1995) attempt to model the
infrared triplet lines of CaII and verify the conditions of their
forming regions. They find that disk accretion, as indicated by the
near IR excess, H$\alpha$ fluxes and veiling, does correlate with the
chromospheric line strengths of CaII and HeI at 5876 ${\rm \AA}$. The
connection is conspicuous and suggests that accretion -- probably
combined with stellar magnetic field lines --  induces regions on the
stellar surface with a distribution of gas density and temperature
similar to those  of solar type magnetically active regions.
Chromospheric lines are therefore enhanced in classical TTS.

In this paper we investigate the photospheres of TTS, largely exploring
the technique of line depth ratios (LDR's). Gray \& Johanson (1991)
devised a method to determine T$_{\rm eff}$ based on the strong
correlation with the LDR of atomic lines, presenting an opposite
response to the increase of T$_{\rm eff}$. The application of this
method to very active stars is not free from drawbacks since the
selective chromospheric filling in and the strong spotting can mask the
actual dependence of line depths with T$_{\rm eff}$. The filling in can
be avoided by a careful selection of the pair of lines among those free
from chromospheric contamination (Finkenzeller \& Basri 1987). The
effect of cool spots on line absorption intensities is negligible for
lines sensitive to temperature such as the LiI resonance line at \lam
6708 ${\rm \AA}$ (Pallavicini et al. 1993) and CaH rotational lines (Barbuy
et al. 1993, Paper I), at least for the typical TTS surface coverage by
spots.  Therefore, we import the LDR technique developed by Gray \&
Johanson (1991), and apply to several TTS (Basri \& Batalha 1990), to
permit preliminary assessment of the intrinsic stellar colors. The
derived stellar parameters are used as an input to synthesize selected
spectral regions. We verify that the LDR of a given pair of lines
presents negligible changes if the spectrum is artificially veiled (0
$<$ {\it v} $<$ 1.0) or rotated (0 $<$ {\it v}sin{\it i} $<$ 20 Km/s)
by the values here reported. Therefore, we adopted LDR's instead of
equivalent width ratios because the former are less sensitive to
spectral noise than the latter.

Since the TTS are undergoing contraction towards the main sequence,
their surface gravities  are expected to be lower than those
characteristic of dwarf stars. In fact, Mould \& Wallis (1977), using
photometry of the 6830 ${\rm \AA}$ CaH band, concluded that TTS have
surface gravities averaging between dwarf and giant values. The
results of Mould \& Wallis have been confirmed since more accurate
values of surface gravity based on improved TTS stellar luminosities
have been determined (see Strom et al. 1989, Cohen, Emerson \& Beichman
1988, Cabrit et al. 1990, Hartmann \& Kenyon 1990 and Hartigan et al.
1991).  In the present work, we explore the use of molecular lines in
high resolution data to better constrain the TTS gravities.

We determine T$_{\rm eff}$ for 13 TTS, most of them belonging to the
Taurus-Aurigae stellar formation complex. We successfuly determine
surface gravities for the cooler stars using spectral synthesis of CaH
and atomic lines. In section 2, we briefly present the observations,
data analysis and T$_{\rm eff}$ determinations based upon published
calibrated main sequence data. In section 3, we describe the synthesis
and the computation procedure to achieve the T$_{\rm eff}$ and
log$\,g$. A brief summary is presented in section 4.
\vskip 1.0 true cm
\noindent {\bf 2. OBSERVATIONS AND DATA ANALYSIS}
\vskip 0.5 true cm
The observations used in this paper result from an extensive
observational program undertaken at Lick Observatory by Dr. Gibor Basri
and collaborators, between 1986 October and 1989 October. This program
includes the observation of the brighter objects in the Taurus-Auriga
complex with the Hamilton-Echelle Spectrograph, aiming to study the
strong emission lines of TTS. The detector used was a TI $800\times
800$ CCD chip. The list of TTS studied in our work is given in Table
1. Some details of the spectral reduction procedure are described by
Basri, Wilcots \& Stout (1989) and Basri \& Batalha (1990).  The
average S/N of the spectra is about 80.

The total echelle region runs from about 5500 to 9000 ${\rm \AA}$ with
gaps in between orders. Those gaps result from the fact that the CCD
used in the observations is smaller than the format of the
Hamilton-Echelle spectrograph, so that there are  cutoffs in the
spectral coverage between consecutive orders.

We select the following spectral sub-regions: \llam 5700-5710 ${\rm
\AA}$ which contains the atomic lines used in the T$_{\rm eff}$
determination; \llam 6770-6820 ${\rm \AA}$ containing some of the most
conspicuous CaH lines found in the sunspot spectrum (Boyer et al. 1975,
1976, 1978 and 1982) and \llam 5400-6400 ${\rm \AA}$ which is used in
the veiling measurements (see below).

We also use spectra of a number of template stars from the Hyades and
the field. These stars are listed in Table 2. They were observed with
the same instrumental apparatus as the TTS. The S/N of the template star
spectra is typically around 200.

Continuum normalization was made using IRAF package routines. The
procedure consists of fitting a spline curve on selected continuum
windows of the given echelle order and then dividing the spectrum of
this order by the fitted curve. In some cases we did not find a unique
solution that  satisfactorily fit the whole order.  In such cases,  we
fit different curves for the intervals $\lambda\lambda 6770-6790 {\rm
\AA}$ and $\lambda\lambda 6790-6810 {\rm \AA}$.

Gravities for three stars are derived by fitting directly CaH molecular
lines.  Therefore, we have  to subtract the continuum nonstellar
contribution from the observed spectrum. The continuum veiling is
expected to be fairly constant redward of 5500 ${\rm \AA}$ (Basri \&
Batalha 1990). Hence, we extract the veiling by applying the simple
expression $$ F_{ph} = (1+v)F_{TT} - v , $$ \noindent where $F_{TT}$ is
the observed TTS profile, relative to a normalized continuum. It is the
sum of the fluxes originating in distinct emitting regions.  $F_{ph}$
is the normalized photospheric flux and $v$ is the continuum veiling,
which is the total non stellar continuum in units of $F_{ph}$.

The $v$-parameter can be determined from its influence on absorption
line depths. Basri \& Batalha (1990) performed such an analysis using
the same spectral database of this work, but in some cases adopting
different spectral types. We adopt their values for every star whose
spectral type coincides with ours. For the other stars, we determine
independent values, comparing the depths of weak unblended atomic lines
between the TTS spectra and those of Hyades templates. The values are
shown in column 7 of Table 1.
\vskip 1.0 true cm
\noindent {\bf 2.1 Calibration of T$_{\rm eff}$}
\vskip 1.0 true cm
Effective temperatures of T Tauri Stars are usually estimated from
their spectral types and the calibration of Cohen \& Kuhi (1979) is the
most widely adopted. Therefore, it is not surprising to find a general
agreement among the temperatures of several CTTS although the
systematic errors generated while translating spectral types to T$_{\rm
eff}$ are not negligible for cool stars (de Jager \& Nieuwenhuijzen
1987, Strom et al. 1988). In this work, we determine the T$_{\rm eff}$
of the program TTS using  the LDR of two absorption lines with
different temperature sensitivities. We first derive intrinsic colors
from which we assess the T$_{\rm eff}$, using a calibration from the
literature.

We carried out a survey of line pairs close in wavelength and free from
strong blends or telluric features. We ended with the following lines:
$\lambda\lambda$ (${\rm \AA}$) ScI-CuI 5700.2/FeI 5706.1 VI 5703.6/FeI
5706.1, VI 5707.0/FeI 5706.0 and FeI 6200.3/ScI 6210.7 (Moore, Minnaert
\& Houtgast 1966). Further analysis led us to exclude some of these
because of the possibility  that they are filled in by chromospheric
emission.  Finally, we selected the pair VI 570.7/570.6, which is shown
in Table 3 together with the atomic parameters. These lines happen to
be the same set previously used by Basri \& Batalha (1990) to infer TTS
spectral types. Their inverse sensitivity to temperature is illustrated
in Figure 1 where we plot the spectra of two Hyades stars with
different spectral types. The line pair of Gray \& Johanson (1991) is
located in one of the gaps of our echelle setting and therefore,  it
could not be included in our study.

We measure the LDR VI 5707/FeI 5706 for a set of field stars and cool
members of the Hyades cluster, with available photometry (see Table
2).  We fit gaussians to the line profiles and take the amplitude of
the curve as the measured line depth. The LDR for each star is given in
Table 2. The (R--I) color indices of these stars are taken from Upgren
(1974) (Hyades) and Gliese (1969) (field stars) and they  are all in
the Kron system. In Figure 2 we present the observed LDR $vs.$
(R--I)$_0$ relation. A larger number of template stars would
statistically improve our relation, especially at the cool end. We
include in section 3.2 the rotational lines of CaH to compute the
fundamental parameters of stars cooler than 4500 K.  A quadratic
polynomial with coefficients ($a_0,a_1,a_2$) = (-0.10901979,
0.58777191, -0.10085137) is then fit to the relation of Figure 2. The
intrinsic T Tauri (R--I)$_0$ colors are then computed using this
polynomial fit and the TTS LDR's.

The TTS LDR's are measured using the same procedure described above.
Given the S/N ratio of our spectra (see Table 1), the average formal
error in the LDR's  due to continuum normalization procedures is equal
to 0.04 for the TTS and 0.01 for the template stars. The average error
induced by the gaussian fitting is equal to 0.06. Therefore, the final
average error in our LDR's is about 0.07 for the TTS and 0.06 for the
templates.

We have observations taken during different seasons for AA, BP, DF and
DK Tau so that we may check for variations of the LDR. The variations
are smaller than the above mentioned error for all stars, except AA
Tau. This will be briefly discussed  in Section 3.2. For the other
stars, we coadded the individual spectra in order to analyze average
data, with improved S/N. The error in the TTS (R--I)$_0$ colors is
about 0.03, based on the mean deviation of the points from the
polynomial fit.

We did not derive colors for DE and DF Tau because their LDR's fall
outside the range encompassed by the set of data assembled in Table 2
(they are much cooler than the template stars). The final color indices
are presented in Table 4 with an average for those stars with more than
one observation. These indices must be considered as first
approximations to the real stellar values since, in our approach, we
are deriving TTS stellar parameters by comparing them to those taken
from a set of main sequence templates. As pointed out by Hartigan et
al. (1991),  the ideal set of templates to be used in a study of CTTS
would be formed by a group of WTTS from the same parental cloud of the
program stars and absent of any signature of disk accretion.  Such
stars, which are also undergoing contraction towards the ZAMS, would
have slightly lower gravities and higher chromospheric activity than
their stable main sequence counterparts. In addition, we would not
expect severe metallicity differences between the template and program
stars, since they are born out from the same interstellar material.

We rely upon the calibration of Soderblom et al. (1993) based on a set
of 87 Pleiades stars to translate colors to effective temperatures.
This cluster is a relatively young one (100 My), and the chromospheric
activity of its stars is comparable to that of a weak T Tauri (Batalha
et al. 1994). We fit a third degree polynomial with coefficientes
($a_0,a_1,a_2,a_3$) = (6943.3565, -6651.5949, 1776.5984, 2290.4328) to
this data, excluding  few stars that deviate more than 3 times the
nearby standard deviation. Effective temperatures of our templates and
TTS are then calculated, and the final values are listed in Tables 2
and 4 respectively. We estimate an error of 100K in the inferred TTS
effective temperatures by measuring the standard deviation of the
residuals around the polynomial fit. However, this is a lower limit to
the error, since systematic effects are not taken into account.
\vskip 0.5 true cm
\noindent {\bf 3. SPECTRUM SYNTHESIS}
\vskip 0.5 true cm
Atomic lines show different responses to changes in gravity. Therefore,
dwarfs and giants of the same temperature might present different
LDR's. The procedure described in the previous section to achieve
effective temperatures will be reliable whenever the TTS is near the
beginning of the ZAMS; otherwise, gravity effects must be taken into
account in the analysis of LDR's. We synthesize the pair of lines from
which the TTS temperatures presented in Table 3 are derived and explore
their response to gravity.  The distinct behaviour of these lines
allows us to make a first guess as to the gravity of individual stars.
We verify these initial guesses of log$\,g$ by synthesizing rotational
lines of CaH.
\vskip 0.5 true cm
\noindent {\bf 3.1 Computation of atomic LDRs and first guess for log$\,g$}
\vskip 0.5 true cm
We have built a theoretical relation between T$_{\rm eff}$ and LDR for
the pair 570.7/570.6, computing synthetic spectra for a grid of model
atmospheres. This grid is computed, with T$_{\rm eff}$ and log$\,g$
ranging from 3600 to 5800K (steps of 200K) and 3.0 to 4.5 (steps of 0.5
dex) and solar metallicity. Each model atmosphere is interpolated in
the grids of Kurucz (1992) and the synthetic spectrum is computed using
the program described in Barbuy (1982). Gray (1994) points out that the
trend of LDR versus T$_{\rm eff}$ might be dependent on metallicity
whenever one of the lines is near saturation.  However, Balachandran \&
Carr (1994) determined solar abundances for some TTS of the
Taurus-Auriga cloud (UX Tau A and TAP 56) which supports our basic
assumption of solar metallicity and the use of our calibration (see
Figure 2) for pre-main-sequence stars.

In order to compute theoretical LDRs, we need the oscillator strengths
({\it gf}) of the involved lines.  In the absence of laboratory
determinations, we derive the {\it gf} values by fitting the spectra of
standard stars. We make several tests before achieving the final {\it
gf} values presented in Table 3.  Both lines of the pair are blends,
which decreases the accuracy of their determined oscillator strengths.
The starting point in our search for the atomic {\it gf} factors is to
fit the disk-integrated solar spectrum of Kurucz et al. (1984) using
two different inputs: the theoretical model of Kurucz (1992) and the
empirical model of Holweger \& M\"uller (1974).  We adopt T$^\odot_{\rm
eff}$ = 5780K and log$\,g^\odot$ = 4.44.  The two sets of {\it gf}
values thus obtained are in perfect agreement.  We also compute another
set of {\it gf} values comparing the theoretical solar spectrum with
the observed solar disk-center spectrum (Delbouille, Roland \& Neven
1973). The set of {\it gf} values thus obtained is substantially
different from the previous one. This is not surprising, since the
solar disk center spectrum does not include the contribution of
limb-darkened layers. Since we are applying these {\it gf} values to
the synthesis of disk integrated stellar spectra, the best suited set
is the one derived from the fit of the Kurucz et al. solar spectrum.

A drawback of semi-empirical {\it gf} values is that of being dependent
on the chosen model atmosphere. Therefore, we verify our previous
figures by matching the spectrum of a well known star, cooler than the
Sun. We compute the synthetic spectrum of Arcturus ($\alpha$ Bootis, HD
124897), interpolating a model atmosphere with T$_{\rm eff}$ = 4340K,
log$\,g$ = 1.6 and [Fe/H] = $-$0.81 (Bell, Edvardsson \& Gustafsson
1985) in the grid of Kurucz model atmospheres. We fit the spectrum of
the photometric atlas of Arcturus (Griffin 1969), but no set of {\it
gf} values satisfies simultaneously the observed lines in the spectrum
of the disk-integrated Sun and that of Arcturus. There are three points
to be considered here.  First, it is possible to use the asymmetries of
the observed solar spectrum profiles in order to guess the {\it gf}
values of individual blended lines which are washed out in the lower
resolution, lower S/N spectrum of Arcturus.   Secondly, the model
atmospheres of Arcturus and the Sun are interpolated in the same grid
and the disagreement may reflect errors in the assumed atmospheric
parameters of Arcturus, since the solar ones are known with the highest
accuracy. Thirdly, the metallicity diference between the Sun and
Arcturus may be large enough that the saturation problems pointed out
by Gray (1994) are important. Thus, we are more confident with the
results based on the solar spectrum.

We use our set of templates as guidelines towards low T$_{\rm eff}$'s
using the effective temperatures derived in Section 2.1. In this way,
we try to match not the observed spectrum of each star, but the
observed trend of the LDR as a function of T$_{\rm eff}$.  In Figure 3,
we plot the relation computed using the solar {\it gf}'s together with
the positions of the Hyades and field stars in the LDR-T$_{\rm eff}$
plane. The location of the templates in Figure 3 lies at about the
trend defined by the computed curves. Hence, the final {\it gf} values,
listed in Table 3, are those which fit the line profiles in the solar
spectrum and followed the expected trend towards low T$_{\rm eff}$'s.
In every calculation we adopt an interaction constant of $C_6=3.0\times
10^{-31}$  for all the lines.

The synthetic spectra range from 5705 to 5709 ${\rm \AA}$ and are convolved
with rotational profiles  to simulate the typical $v$sin$i$  of the
program stars. The LDR's do not vary significantly  for $v$sin$i$
between 5 and 20 Km/s. We measure the depths of the lines in each
theoretical spectrum and compute the final relations shown in Figure 3
for log$\,g$ = 3.0, 3.5, 4.0 and 4.5. In Table 5, we present several
LDRs and their associated temperatures and gravities. The strong
dependence of the LDR on log$\,g$ at the cool end of Figure 3 is
striking. This is because iron is predominantly neutral at low
temperatures while vanadium is still partially ionized. Therefore, the
intensity of the neutral iron line is a function of electronic
pressure, while the intensity of the neutral vanadium line is not (Gray
1976, chap. 12), yielding, as a result, different responses to gravity
at low temperatures. As the temperature increases, both elements become
equally ionized, so that the intensity of the neutral lines stops being
a function of eletronic pressure and the LDR is thus insensitive to
gravity.
\vskip 1.0 true cm
\noindent {\bf 3.2 Computation of the CaH lines: Temperature and Gravity
Determinations}
\vskip 0.5 true cm
The model computations of Section 3.1 show that the relation between
the LDR and T$_{\rm eff}$ is degenerate, in the sense that, for one
given LDR, we can have a number of T$_{\rm eff}$'s, depending on the
gravity of the star (Figure 3). This effect is more important for stars
cooler than 4500K. In order to remove this degeneracy, we need one more
relation between LDR, T$_{\rm eff}$ and log$\,g$. In Section 2, we used
an empirical relation between LDR and T$_{\rm eff}$, which can only be
applied to dwarf stars. In this Section, we explore the dependence of
CaH lines on T$_{\rm eff}$ and log$\,g$, with the aim of breaking this
degeneracy.

CaH lines are present in the wavelength range $\lambda\lambda$ 6100 $-$
7100 ${\rm \AA}$. The $\lambda\lambda$ 6770 $-$ 6810 ${\rm \AA}$ region
ended up to be the best suited for studies of CaH lines because of the
weakness of CN and TiO molecular lines and the relatively low number of
blended atomic lines. The region around 6390 ${\rm \AA}$ could also be
used, but, unfortunately, it falls within one of the gaps in the
spectral coverage.

The spectral synthesis code and atomic plus molecular data used are
described in Paper I, in which the CaH B$\sp 2$$\Sigma$ $-$ X$\sp
2$$\Sigma$ and A$\sp 2$$\Pi$ $-$ X$\sp 2$$\Sigma$ systems are studied.
In that paper, the oscillator strengths for the two electronic
transitions are chosen so that we could reproduce, on average, the
observed high-resolution spectrum of a number of Hyades and field
stars.  Here, we redetermine these oscillator strengths, by fitting in
a more rigorous way the spectra of two stars from the Hyades cluster,
VA 404 and VA 622. The temperatures adopted here for these stars are
more reliable than the values of Paper I. The T$_{\rm eff}$'s are taken
from the Kron (R$-$I)$_0$ colors, as explained in Section 2.1 (Table
4), while the temperatures adopted in Paper I are derived from spectral
types. The gravities are taken from the positions of these stars in
Figure 2 (the two coolest stars); for VA 404, we have log$\,g$ = 4.0
and for VA 622, log$\,g$ = 4.5. The adopted  model atmospheres are
interpolated in the grids from Kurucz (1992) assuming solar
metallicities. We derive $f_{el}^{A-X} = 1.2$ (0.5 in Paper I) and
$f_{el}^{B-X} = 1.4$ from this analysis. The latter is less reliable,
since it depends upon the fitting of fewer lines than the former. In
Figure 4, we display the response of CaH lines to temperature, (for
log$\,g$ = 4.0) and gravity  (for T$_{\rm eff}$ = 4000 K).

Atomic neutral lines usually grow stronger towards low gravities
because of the larger concentration of neutral metals at the same
optical depth.  Hydride lines behave in the opposite manner, because
the formation of the ionic molecule tends to increase with increasing
gas pressure as the star approaches the main sequence. Therefore, the
(veiling independent) LDR between one CaH line and one atomic line is
used here to further constrain the stellar parameters. If we model the
dependence of this new LDR upon the stellar parameters, we will have
the additional relation needed to break the degeneracy apparent in
Figure 3. We synthesize the region between 6790 and 6810 ${\rm \AA}$
and the best CaH line is chosen.  We use the line at 6796 ${\rm \AA}$,
which shows a good response to log$\,g$, is fairly deep and easily
found in our set of CTTS spectra.  Next, we compute two new LDR's, CaH
6796/FeI 5706 and CaH 6796/VI 5707, for the grid of model atmospheres
referred to in Section 3.1.  This gives us the theoretical dependence
of the LDRs upon T$_{\rm eff}$ and log$\,g$ (for solar metallicity).
(See also the discussion of the VI 5707/FeI 5706 LDR in Section 3.1).
The responses of the two new LDR's to T$_{\rm eff}$ and log$\,g$ are
very similar to each other. One of these relations is shown in Figure
5.

We use each one of these two LDR's to form a pair with the LDR of
Section 3.1 and use the model relations of Figures 3 and 5 to search
for a set of T$_{\rm eff}$ and log$\,g$ which consistently fit the
measurements of each of the two LDR's thus formed ({\it e.g.}, [VI/FeI,
CaH/FeI] and [VI/FeI, CaH/VI]). For each pair of LDR's, a solution is
found whenever the T$_{\rm eff}$'s and log$\,g$'s indicated by the
LDR's disagree by less than 100K and 0.4$dex$, respectively. We then
take the average of the values given by each LDR.  Hence, we have two
determinations of T$_{\rm eff}$ and log$\,g$ for each star: one
obtained from the pair [VI/FeI, CaH/FeI] and the other from the pair
[VI/FeI, CaH/VI]. We consider the solutions falling inside the range
[3600,4400K], [2.5,4.5 $dex$] to be reliable. We find that the [VI/FeI,
CaH/FeI] pair converges better than the other pair, in most cases, to a
unique solution ({\it i.e.}T$_{\rm eff}$'s and log$\,g$'s as indicated
by each LDR differing by less than 50K and 0.1$dex$, respectively).
This could be due to the fact the for stars cooler than 3900K, the VI
line begins to saturate, which lowers the sensitivity of the CaH/VI LDR
to both log$\,g$ and T$_{\rm eff}$.

We have observations on more than one night for AA, BP, DF and DK Tau.
We can therefore check the stability of the LDR's against variations
which may occur on timescales as short as one day.  As mentioned in
Section 2, the variations of LDR's in the spectra of AA Tau can not be
accounted for by measurement errors alone. In fact, they are loosely
correlated with veiling in the sense that data with larger veilings
show ``hotter'' LDR's. Therefore, we choose the data with the best S/N
and the lowest veiling in order to minimize the effect of any possible
interference of disk accretion with the photospheric lines (see Batalha
et al. 1995).

The final results are shown in Table 4 with superscripts $syn$.  Both
pairs yield the same result for AA Tau, BP Tau and CI Tau. The results
of DE Tau, DF Tau and DK Tau are based solely on the [VI/FeI, CaH/FeI]
pair. For V830 Tau, a consistent, but different, set of solutions
converge for each LDR pair. Both are listed in Table 4. The
uncertainties for this star are larger, given the low S/N of the
spectrum and the relatively high $v$sin$i$ ($\sim 25 km/s$). The
results are displayed as points on Figure 5. There is a fair agreement
between the T$_{\rm eff}$'s determined by the LDR-pair method and those
inferred from the intrinsic (R--I) colors, mainly for the stars with
log$\,g\,\,>$ 4.0. This is because we used a calibration based on dwarf
stars in order to derive the intrinsic colors of the TTS. That
calibration is expected to provide bluer colors   for stars with lower
gravities, because of the gravity dependence of the atomic LDR used
(Figure 3). Indeed, for AA, CI and V830 Tau (log$\,g\,\sim$ 3.8 $dex$),
the T$_{\rm eff}^c$'s exceed T$_{\rm eff}^{syn}$'s by $\sim$ 100K as
shown in Table 4.

We do not find a set of atmospheric parameters satisfying the LDR
measurements for ROX 6, GM Aur and DN Tau. We therefore compute
log$\,g$ for these stars by directly comparing the observed CaH lines
with the synthetic ones, after proper veiling corrections are carried
out (see Section 2).  The synthetic spectra are computed following the
procedure already described, and the adopted T$_{\rm eff}$'s are
derived from the intrinsic (R-I) colors. The chosen log$\,g$ for each
star is the one that gives the best fit to the observed CaH lines. The
uncertainty involved in this method is larger, given the very high S/N
it requires, and given the errors in the veiling and in the adopted
$v$sin$i$. ROX 6 has the largest uncertainty due to the poor S/N of its
spectrum.

Because of the incompleteness of our atomic line list and the errors in
the atomic oscillator strengths, the whole $\lambda\lambda$ 677.0-681.0
region is not well fitted by the synthetic spectra. Thus, we restrict
the analysis to five sub-regions where blending is minimal, i.e., where
Moore, Minnaert \& Houtgast (1966) report no atomic lines. The fits are
shown in Figure 6 and the  gravities thus determined are shown in Table
4.
\vskip 1.0 true cm
\noindent {\bf 3.3 Comparing Previous Determinations of log$\,g$}
\vskip 0.5 true cm
In order to compare our results to the data available in the
literature, we compute surface gravities using luminosity and T$_{\rm
eff}$ data from a number of authors. We locate each star on the H-R
diagram and derive its gravity by interpolating among the theoretical
evolutionary tracks from Cohen \& Kuhi (1979). Spectral types (or
temperatures, when available) and luminosities are from (Cohen \& Kuhi
1979, Cohen, Emerson \& Beichman 1988; Walter et al. 1988; Strom et
al.  1989; Cabrit et al. 1990; Hartigan et al. 1991; Basri Martin \&
Bertout 1991, Magazz\`u, Rebolo \& Pavlenko 1992). For those papers
which only provide spectral types, we estimate the temperatures using
the calibration of de Jager \& Nieuwenhuijzen (1987). The resulting
log$\,g$'s are listed in Table 6. The spread among the values from
various sources for some stars is quite striking.  This is most likely
due to the differing methods used to remove veiling (when done ) and to
correct for reddening. We also show the gravities obtained in the same
way for the stars hotter than 4500K -- T Tau, TAP 56 and UX Tau A --
for which the CaH lines are too faint to provide any determination and
the temperatures were estimated from the calibration of Soderblom et
al. (1993).

The agreement of our gravities with the ones from the literature is
fairly good for AA, CI and V830 Tau. Our values tend to be above the
average for the other stars. These differences can be partially
explained in terms of the differences in the adopted temperatures. In
order to test this hypothesis, we recompute the log$\,g$ for these
stars by placing them again in the HR diagram using our T$_{\rm eff}$'s
rather than the ones derived from the spectral types. The results are
shown in Table 6, inside parentheses, for the stars whose T$_{\rm
eff}$'s as determined by us and other authors differ by 100K or more.
We can see that, at least for DK Tau, DN Tau and GM Aur, the values
inside parentheses are in good agreement with our log$\,g$
determinations.  We conclude that, for these  stars, differences in
log$\,g$ can be reconciled by differences in assumed stellar
temperatures.

For ROX 6, however, a large discrepancy remains -- large enough that it
cannot be explained even by luminosity errors in the work of Cohen \&
Kuhi (1979). As can be seen in Figure 6, there is significant noise in
the CaH lines which implies that we are not confidently determining
either the CaH line depth or the line shape, both basic to the
successful aplication of the LDR method or direct synthesis.  The cases
of DE Tau and DF Tau are even more drastic, with discrepancies in
log$\,g$ of nearly 1 $dex$.  It is likely that  our values are
overestimated since these stellar luminosities have been carefully
measured (Strom et. al 1989) and they imply gravities lower than those
we derive. As discussed previously, errors in the atomic and molecular
data as well as in the adopted opacity sample are inherent in the
theoretical LDR's and they certainly become larger at low
temperatures.  Furthermore, the contrast of the iron line at 5706 ${\rm
\AA}$ against the nearby continuum is severely decreased at
temperatures lower than 3800 K. Thus, the line becomes inceasingly
washed out by noise, especially in DE and DF Tau whose lines are
already sizably veiled.

We conclude that the combined use of CaH with atomic lines to assess
gravities and temperatures of TTS  is efficient for temperatures in the
range 4400 K -- 3800 K.  Table 5 provides a quick and easy assessment
of the fundamental parameters of TTS with spectral types between K5 --
M0. We foresee several improvements in our method such as:  {\it i)}
enlargement of the present set of line pairs, including others more
efficient at temperatures lower than 3800 K; {\it ii)} inclusion of
other hydride lines such as MgH in order to better constrain log$\,g$.
{\it iii)} Finally, future synthetic spectra will be more reliable if
better atomic and molecular oscillator strengths, calibrated with a
selected set of standards stars, are employed.
\vskip 1.0 true cm
{\bf 4. Summary}
\vskip 0.5 true cm
There is a growing need for spectral classification based on high
resolution data, especially for young main sequence stars.  We have
derived temperatures and gravities of a small set of T Tauri stars,
exploring the line depth ratio (LDR) technique, complemented by a
spectral synthesis analysis of atomic and CaH rotational lines.

Temperatures are derived using an empirical relation between (R--I)$_0$
color indices and selected atomic LDRs of a set of template stars in
the local field and the Hyades open cluster. Usually, TTS temperatures
are derived from spectral types which are largely based on low
dispersion spectra.  Such analyses are subject to error due to the
effects of veiling, metallicity, and the conversion between spectral
type and T$_{\rm eff}$.  Our method is significantly less sensitive to
veiling and metallicity, and therefore, produces more reliable
temperature determinations.

At temperatures lower than 4400 K, we find a strong dependence of the
LDR on gravity. The CaH lines are therefore included in the analysis in
order to constrain log$\,g$.  This is accomplished by taking the ratio
of the line depth of a rotational CaH line and two atomic lines.
Computing theoretical relations between LDRs and T$_{\rm eff}$ and
log$\,g$ we derive a best value for each of these parameters.

For some stars, we are not able to find a pair of  T$_{\rm eff}$ and
log$\,g$ which satisfies the LDR measurements. Their gravities are then
obtained by fitting the observed CaH rotational lines with synthetic
lines, adopting the T$_{\rm eff}$ derived from the intrinsic (R--I)$_0$
colors. The gravities derived for stars cooler than 3800K may be
overestimated. The error can be mostly due to the low contrast between
the iron line (5706 ${\rm \AA}$) and the local continuum.  We conclude
that the atomic line pair FeI/VI (5706/5707 ${\rm \AA}$) combined with
the stronger CaH line here studied (6796 ${\rm \AA}$)  yields an
efficient and quick method to assess effective temperature and
log$\,g$ for TTS within the range 3800 $<$ T$_{\rm eff}$ $<$ 4400K,
being completely independent of veiling. We present the ne cessary
calibration to derive these fundamental stellar parameters (Table 5)
given the observed line depths.

In future work, additional observations of a larger number of standard
stars and TTS with higher S/N are desirable in order to define new sets
of LDRs and extend the method towards stars  cooler than a typical M0
star. Likewise, it would be desirable to explore the rotational lines
of MgH to confirm the values here derived and to refine the method. A
study is being undertaken in order to calibrate the intensity of the
near infrared FeH bands as a function of stellar parameters (Piorno
Schiavon, Barbuy \& Singh 1995, in preparation). Those bands are
important gravity indicators for M-dwarf stars.
\vskip 0.5 true cm
\noindent{\it Acknowledgements.} We thank Gibor Basri for making several
of his spectra available. We also thank Dr. D.F. Gray for helpful
discussions on the line depth method and E.L. Martin, N. Stout-Batalha
and the anonymous referee for their comments on the manuscript. R.P.S.
would like to thank G.F. Porto de Mello for encouraging discussions and
CAPES and FAPESP, for financial support.
\vskip 1.0 true cm
\noindent {\bf References}
\vskip 0.5
true cm
\ref Balachandran, S. \& Carr, J.: 1994, in ``8$^{th}$ Cambridge Workshop on
Cool Stars, Stellar Systems and the Sun'', ed. J.-P. Caillaut.

\ref Barbuy, B.: 1981, A\&A, 101, 365.

\ref Barbuy, B.: 1982, PhD. Thesis, Universit\'e de Paris, VII.

\ref Barbuy, B., Schiavon, R.P., Gregorio-Hetem, J., Singh, P.D. and
Batalha, C.: 1993, A\&AS, 101, 409 (Paper I)

\ref Basri, G. \& Batalha, C.: 1990, ApJ, 363, 654.

\ref Basri, G., Wilcots, E. and Stout, N.: 1989, PASP, 101,
528.

\ref Basri, G., Martin, E.L., Bertout, C.: 1991, A\&A, 252, 625
(BMB91)

\ref Batalha, C.C. \& Basri, G.: 1993, ApJ, 412, 363.

\ref Batalha, C.C. et al.:1995 in preparation

\ref Bell, R.A., Edvardsson, B. \& Gustafsson, B.: 1985, MNRAS, 212,
497.

\ref Bertout, C.: 1989, ARA\&A, 27, 35.

\ref Bertout, C., Basri, G., Bouvier, J.: 1988, ApJ, 330, 373

\ref Bouvier, J.: 1990, AJ, 99, 946.

\ref Boyer, R., Sotirovski, P. and Harvey, J.W.: 1975, A\&AS, 19, 359.

\ref Boyer, R., Sotirovski, P. and Harvey, J.W.: 1976, A\&AS, 24, 111.

\ref Boyer, R., Sotirovski, P. and Harvey, J.W.: 1978, A\&AS, 33, 145.

\ref Boyer, R., Sotirovski, P. and Harvey, J.W.: 1982, A\&AS, 47, 145.

\ref Cabrit, S., Edwards, S., Strom, S.E., Strom, K.M.: 1990,
ApJ, 354, 687. (CESS)

\ref Calvet, N., Basri, G. \& Kuhi, L.V.: 1984, ApJ, 277, 725.

\ref Calvet, N. \& Mar\i\ n, Z.: 1987, Rev. Mex. Astr. Ap., 14, 353.

\ref Cayrel de Strobel G., Hauck, B., Fran\c cois, P., Th\'evenin, F.,
Friel, E.  Mermilliod, M., Borde, S.: 1992, A\&AS, 95, 273.

\ref Cohen, M., Emerson, J.P., Beichman, C.A.: 1988, ApJ, 339, 455
(CEB88)

\ref Cohen, M., Kuhi, L.V.: 1979, ApJS, 41, 743 (CK79)

\ref de Jager, C., Nieuwenhuijzen, H.: 1987, A\&A, 177, 217

\ref Delbouille, L., Roland, G., Neven, L.: 1973, ``Photometric
Atlas of the Solar Spectrum from 3000 to 10000 \a'', Institut
d'Astrophysique de Li\`ege

\ref Edwards, S., Hartigan, P., Shandon, L., Andrules: 1994, ApJ, accepted.

\ref Finkenzeller, U., Basri, G.: 1987, ApJ, 318, 823.

\ref Franchini, M., Castelli, F. \& Stalio, R.: 1991, A\&A, 241, 449.

\ref Gliese, W.:  1969, Ver\"off. Astron. Rech. Inst., Heidelberg, 22.

\ref Gray, D.F.: 1994, private comunication

\ref Gray, D.F., Johanson, H.L.: 1991, PASP, 103, 439.

\ref Griffin, R.F.: 1969, ``A photometric atlas of the spectrum of
Arcturus $\lambda\lambda 3600 - 8825 \AA$'', Cambridge Philosophical
Society.

\ref Hartigan, P., Hartmann, L., Kenyon, S.J., Hewett, R. and Stauffer,
J.: 1989, ApJ, 70, 899.

\ref Hartigan, P., Hartmann, L., Kenyon, S.J., Strom, S.E., Skrutskie,
M.F.: 1990, ApJ, 354, L25.

\ref Hartigan, P., Kenyon, S.J., Hartmann, L., Strom, S.E., Edwards,
S., Welty, A.D. and Stauffer, J.: 1991, ApJ, 382, 617.
(HKHSEWS).

\ref Hartmann, L.W. \& Kenyon, S.J.: 1987, ApJ, 349, 243.

\ref Hartmann, L.W. \& Kenyon, S.J.: 1990, ApJ, 349, 190.

\ref Herbig, G.H. \& Bell, K.R.: 1988, Lick Obs. Bull. 1111.

\ref Holweger, H., M\"uller, E.: 1974, Solar Phys. 39, 19.

\ref Kenyon, S.J., Hartmann, L. Hewett R., Carrasco, L. Cruz-Gonzalez, I.,
Recillas, E., Salas, L. \& Serrano, A.: 1994, AJ, 107, 2153.

\ref K\"onigl, A.: 1991, ApJ, 370, L39.

\ref Kurucz, R.L., Furenlid,I., Brault,J., Testerman,L.: 1984,
``Solar Flux Atlas from 296 to 1300 nm ''(Sunspot, N.M., National Solar
Observatory).

\ref Kurucz, R.: 1992, in IAU Symp. 149, 225

\ref Magazz\`u, A., Rebolo, R., Pavlenko, Y.V.: 1992, ApJ, 392, 159. (MRP92)

\ref Martin, E.L., Rebolo, R., Magazz\`u, A. \& Pavlenko, Y.V.: 1994, A\&A,
282, 503.

\ref Moore, C.E., Minnaert, M.G.J. and Houtgast, J.:  1966, The
``Solar Spectrum 2935\a\ to 8770\a'' (Washington: National Bureau of
Standards) (NBS Monograph 61).

\ref Mould, J.R. \& Wallis, R. E.: 1977, MNRAS, 181, 625.

\ref Pallavicini, R., Cutispoto, G., Randich, S., Gratton, R.: 1993,
A\&A, 267, 145

\ref Shu, F., Najita, J., Ostriker, E., Wilkin, F., Ruden, S.
\& Lizano, S.: 1994, ApJ, 429, 781.

\ref Soderblom, D.R., Stauffer,J.R., Daniel Hudon, J. \& Jones, B.F.:
1993, ApJS, 85, 315.

\ref Strom, K.M., Strom, S.E., Kenyon, S.J. and Hartmann, L.: 1988,
AJ, 95, 534.

\ref Strom, K.M., Strom, S.E., Edwards, S., Cabrit, S., Skrutskie,
M.F.:  1989, AJ, 97, 1451 (SSECS)

\ref Upgren, A.R. .: 1974, ApJ, 193, 359.

\ref Walter, F.M., Brown, A., Mathieu, R.D., Myers, P.C., Vrba, F.J.: 1988,
AJ, 96, 297 (WBMMV)

\vfill\eject

\noindent{\bf Table 1} - Basic data for program stars

$$\vbox{\halign to \hsize{\tabskip 1em plus2em$
\quad\quad \hfil #\hfil$&
$\hfil  #\hfil$&
 $\hfil #\hfil\quad$&
 $\hfil #\hfil$&
$\hfil # \hfil $\cr
\noalign{\hrule}
\noalign{\vskip 0.2cm}
{\rm Star} & {\rm Date}  & {\rm SpT} & {\rm
v{\rm sin}i (km/s)} & {\rm veiling} \cr
\noalign{\vskip 0.1cm\hrule\vskip 0.4cm}
{\rm AA \,\,Tau} & 12/22/86 &{\rm K9} & 10 & 0.1 \cr
\noalign{\vskip 0.15cm}
{\rm BP \,\,Tau} & average & {\rm K7} & < 10 & \cr
\noalign{\vskip 0.15cm}
{\rm CI \,\,Tau} & 10/12/87  & {\rm K6} & 11 & 0.1 \cr
\noalign{\vskip 0.15cm}
{\rm DE \,\,Tau} & 10/12/87 & {\rm M2} & < 10 & 0.9 \cr
\noalign{\vskip 0.15cm}
{\rm DF \,\,Tau} & average  & {\rm K7} & 10 & \cr
\noalign{\vskip 0.15cm}
{\rm DK \,\,Tau} & average  & {\rm M0} & 10 & \cr
\noalign{\vskip 0.15cm}
{\rm DN \,\,Tau} & 11/12/86  & {\rm K7} & 10 & 0.0 \cr
\noalign{\vskip 0.15cm}
{\rm GM \,\,Aur} & 10/11/87  & {\rm K4} & 13 & 0.2 \cr
\noalign{\vskip 0.15cm}
{\rm ROX \,\,6} & 09/06/87  & {\rm K5} & < 10 & 0.3 \cr
\noalign{\vskip 0.15cm}
{\rm T \,\,Tau} & 10/11/87  & {\rm K2} & 20 & 0.1 \cr
\noalign{\vskip 0.15cm}
{\rm TAP \,\,56} & 11/29/88   & {\rm K2} & 22 & 0.0 \cr
\noalign{\vskip 0.15 cm}
{\rm UX \,\,Tau \,\,A} & 12/22/86  & {\rm K2} & 20 & 0.2 \cr
\noalign{\vskip 0.15cm}
{\rm V830 \,\,Tau} & 11/12/86 & {\rm K6} & 25 & 0.0 \cr
\noalign{\vskip 0.15cm}
\noalign{\vskip 0.30cm}  \cr}
\vskip 0.3 cm
\hrule}$$

\vfill\eject

\noindent{\bf Table 2} - Basic data for template Hyades and field stars:

$$\vbox{\halign to \hsize{\tabskip 1em plus2em$
#\hfil$&
$ #\hfil $ &
 $ #\hfil$&
 $ #\hfil$&
 $ #\hfil$&
 $ #\hfil$&
$# \hfil $\cr
\noalign{\hrule}
\noalign{\vskip 0.2cm}
{\rm Star} & {\rm Date} & {\rm V} & {\rm SpT} & {\rm line\,\,ratio} & {\rm
(R-I)_0}
& {\rm T_{eff} (K)} \cr
\noalign{\vskip 0.1cm\hrule\vskip 0.4cm}
{\rm HR 1262} & 8/12/89 & 5.90  & {\rm G5} & 0.48 & & \cr
\noalign{\vskip 0.15cm}
{\rm HR 159} & 8/12/89 & 5.57 & {\rm G8} & 0.60 & & \cr
\noalign{\vskip 0.15cm}
{\rm HR 8} & 8/12/89 & 6.13 & {\rm K0} & 0.64 & & \cr
\noalign{\vskip 0.15cm}
{\rm HR 493} & 8/12/89 & 5.20 & {\rm K1} & 0.83 & 0.29 & 5220 \cr
\noalign{\vskip 0.15cm}
{\rm HR 222} & 8/12/89 & 5.75 & {\rm K2} & 0.92 & 0.33 & 5020 \cr
\noalign{\vskip 0.15cm}
{\rm HR 753} & 8/12/89 & 5.83 & {\rm K3} & 1.13 & 0.36 & 4890 \cr
\noalign{\vskip 0.15cm}
{\rm VA 587} & 14/10/89 & 8.93 & {\rm K0.5} & 0.76 & 0.31 & 5120 \cr
\noalign{\vskip 0.15cm}
{\rm VA 459} & 8/12/89 & 9.45 & {\rm K2} & 0.87 & 0.34 & 4980 \cr
\noalign{\vskip 0.15cm}
{\rm VA 135} & 14/10/89 & 9.99 & {\rm K4} & 1.14 & 0.47 & 4450 \cr
\noalign{\vskip 0.15cm}
{\rm VA 276} & 8/12/89 & 10.52 & {\rm K5} & 1.32 & 0.49 & 4380 \cr
\noalign{\vskip 0.15cm}
{\rm VA 404} & 14/10/89 & 10.51 & {\rm K7} & 1.71 & 0.63 & 4030 \cr
\noalign{\vskip 0.15cm}
{\rm VA 622} & 14/10/89 & 11.96 & {\rm M0} & 2.23 & 0.69 & 3950 \cr
\noalign{\vskip 0.30cm}  \cr}
\vskip 0.3 cm
\hrule}$$

\vfill\eject

\noindent {\bf Table 3} - Line pair used in the temperature determination.

$$\vbox{\halign to \hsize{\tabskip 1em plus2em$
\quad\quad \hfil #\hfil$&
$\hfil  #\hfil$&
 $\hfil #\hfil\quad$&
$\hfil #\hfil $ &
 $\hfil #\hfil$&
$\hfil # \hfil $$\cr
\noalign{\hrule}
\noalign{\vskip 0.2cm}
{\rm Species} & \lambda ($nm$) & {\rm \chi_{ex}\; (eV)} & {\rm log \; gf}
&\cr
\noalign{\vskip 0.15cm}  \cr
\noalign{\hrule}
\noalign{\vskip 0.15cm}  \cr
{\rm Fe1} & 570.601 & 4.61 & -0.80 & \cr
{\rm Fe1} & 570.611 & 4.28 & -1.95 & \cr
{\rm V1} & 570.698 & 1.04 & -0.6 & \cr
{\rm Fe1} & 570.705 & 3.64 & -2.48 & \cr
\noalign{\vskip 0.1cm}  \cr}
\vskip 0.3 cm
\hrule}$$

\vfill\eject

\noindent {\bf Table 4 } - Line ratios, temperatures and gravities.
$$\vbox{\halign to \hsize{\tabskip 1em plus2em
$\quad \hfil #\hfil$&
$\hfil #\hfil$&
$\hfil #\hfil$&
$\hfil #\hfil$&
$\hfil #\hfil$&
$\hfil #\hfil \quad $\cr
\noalign{\hrule\vskip 0.4cm}
{\rm Name}
 & \hidewidth {\rm line\,\, ratio} \hidewidth
 & \hidewidth {\rm (R-I)_0} \hidewidth
& \hidewidth {\rm T_{\rm eff}^{\rm c}}\hidewidth
& \hidewidth  {\rm T_{\rm eff}^{\rm syn}} \hidewidth
& \hidewidth {\rm  log\;{\it g}}^{\rm syn}\hidewidth \cr
\noalign{\vskip 0.4cm\hrule\vskip 0.4cm}
{\rm AA\;Tau} & 2.02 & 0.67 & 3980 & 3840 & 3.9 \cr
\noalign{\vskip 0.15cm}
{\rm BP\;Tau} & 1.77 & 0.63 & 4060 & 4100 & 4.3 \cr
\noalign{\vskip 0.15cm}
{\rm CI\;Tau} & 1.42 & 0.54 & 4270 & 4200 & 3.8 \cr
\noalign{\vskip 0.15cm}
{\rm DE\;Tau} & 3.04 &  & & 3690 & 4.2 \cr
\noalign{\vskip 0.15cm}
{\rm DF\;Tau} & 2.98 &  & & 3660 & 4.0 \cr
\noalign{\vskip 0.15cm}
{\rm DK\;Tau} & 1.39 & 0.51 & 4320 & 4260 & 4.2 \cr
\noalign{\vskip 0.15cm}
{\rm DN\;Tau} & 1.68 & 0.61 & 4100 & & 3.8^\dagger \cr
\noalign{\vskip 0.15cm}
{\rm GM\;Aur} & 1.18 & 0.47 & 4490 & & 4.5^\dagger \cr
\noalign{\vskip 0.15cm}
{\rm ROX\;6} & 1.34 & 0.52 & 4330 & & 4.2^\dagger \cr
\noalign{\vskip 0.15cm}
{\rm T\;Tau} & 0.95 & 0.38 & 4870 & &  \cr
\noalign{\vskip 0.15cm}
{\rm TAP\;56} & 0.96 & 0.39 & 4860 & &  \cr
\noalign{\vskip 0.15cm}
{\rm UX\;Tau\;A} & 0.93 & 0.38 & 4920 & &  \cr
\noalign{\vskip 0.15cm}
{\rm V830\;Tau} & 1.45 & 0.55 & 4240 & 4010-4160 & 3.6-4.2 \cr
\noalign {\vskip 0.3 cm} \cr}
\hrule}$$

Those values indicated by $\dagger$ were obtained from the
synthesis of CaH lines. The others were obtained by interpolating the
model relations between line depth ratios and [T$_{\rm eff}$,
log$\,g$].

\baselineskip 12 pt

\noindent {\bf Table 5} - Line depth ratios computed from the synthetic
spectra.
$$\vbox{\halign to \hsize{\tabskip 1em plus2em
$\quad \hfil #\hfil$&
$\hfil #\hfil$&
$\hfil #\hfil$&
$\hfil #\hfil$&
$\hfil #\hfil$&
$\hfil #\hfil$&
$\hfil #\hfil$&
$\hfil #\hfil$&
$\hfil #\hfil\quad $$\cr
\noalign{\hrule\vskip 0.2cm}
& {\rm V/Fe} & {\rm CaH/Fe} & {\rm CaH/V} & & {\rm V/Fe} & {\rm CaH/Fe} & {\rm
CaH/V} \cr
\noalign{\vskip 0.2cm\hrule\vskip 0.3cm}
& & {\rm log}\,g\,=\,3.0 & & & & {\rm log}\,g\,=\,4.0  & \cr
{\rm T_{eff} (K)} & & & & {\rm T_{eff}(K)} & & & \cr
\noalign{\vskip 0.2 true cm}
3600 & 2.04 & 1.94 & 0.81 & 3600 & 3.19 & 3.56 & 0.98 \cr
3800 & 1.68 & 0.95 & 0.51 & 3800 & 2.28 & 1.61 & 0.67 \cr
3900 &      & 0.65 & 0.39 & 3900 &      & 1.16 & 0.56 \cr
4000 & 1.44 & 0.50 & 0.32 & 4000 & 1.73 & 0.90 & 0.50 \cr
4100 &      & 0.32 & 0.22 & 4100 &      & 0.71 & 0.42 \cr
4200 & 1.28 & 0.19 & 0.14 & 4200 & 1.46 & 0.52 & 0.34 \cr
4300 &      & 0.12 & 0.09 & 4300 &      & 0.31 & 0.23 \cr
4400 & 1.18 & 0.06 & 0.05 & 4400 & 1.29 & 0.22 & 0.17 \cr
4500 &      & 0.02 & 0.02 & 4500 &      & 0.15 & 0.12 \cr
4600 & 1.08 &      &      & 4600 & 1.14 &      &      \cr
4800 & 0.99 &      &      & 4800 & 1.02 &      &      \cr
5000 & 0.91 &      &      & 5000 & 0.92 &      &      \cr
5200 & 0.82 &      &      & 5200 & 0.82 &      &      \cr
5400 & 0.72 &      &      & 5400 & 0.71 &      &      \cr
5600 & 0.63 &      &      & 5600 & 0.62 &      &      \cr
5800 & 0.54 &      &      & 5800 & 0.53 &      &      \cr
\noalign{\vskip 0.3cm\hrule\vskip 0.3cm}
& & {\rm log}\,g\,=\,3.5 & & & & {\rm log}\,g\,=\,4.5  & \cr
{\rm T_{eff} (K)} & & & & {\rm T_{eff}(K)} & & & \cr
\noalign{\vskip 0.2 true cm}
3600 & 2.45 & 2.56 & 0.91 & 3600 & 4.46 & 5.27 & 1.09 \cr
3800 & 1.88 & 1.28 & 0.60 & 3800 & 2.99 & 2.15 & 0.72 \cr
3900 &      & 0.90 & 0.48 & 3900 &      & 1.55 & 0.62 \cr
4000 & 1.56 & 0.71 & 0.41 & 4000 & 2.11 & 1.24 & 0.58 \cr
4100 &      & 0.50 & 0.32 & 4100 &      & 0.93 & 0.50 \cr
4200 & 1.36 & 0.33 & 0.23 & 4200 & 1.65 & 0.74 & 0.44 \cr
4300 &      & 0.19 & 0.15 & 4300 &      & 0.53 & 0.35 \cr
4400 & 1.22 & 0.11 & 0.08 & 4400 & 1.39 & 0.38 & 0.28 \cr
4500 &      & 0.06 & 0.05 & 4500 &      & 0.24 & 0.20 \cr
4600 & 1.10 &      &      & 4600 & 1.19 &      &      \cr
4800 & 1.00 &      &      & 4800 & 1.03 &      &      \cr
5000 & 0.92 &      &      & 5000 & 0.92 &      &      \cr
5200 & 0.82 &      &      & 5200 & 0.81 &      &      \cr
5400 & 0.72 &      &      & 5400 & 0.69 &      &      \cr
5600 & 0.63 &      &      & 5600 & 0.60 &      &      \cr
5800 & 0.54 &      &      & 5800 & 0.52 &      &      \cr
\noalign{\vskip 0.4cm\hrule\vskip 0.2cm}}}$$

\vfill
\eject

\tabmongofig{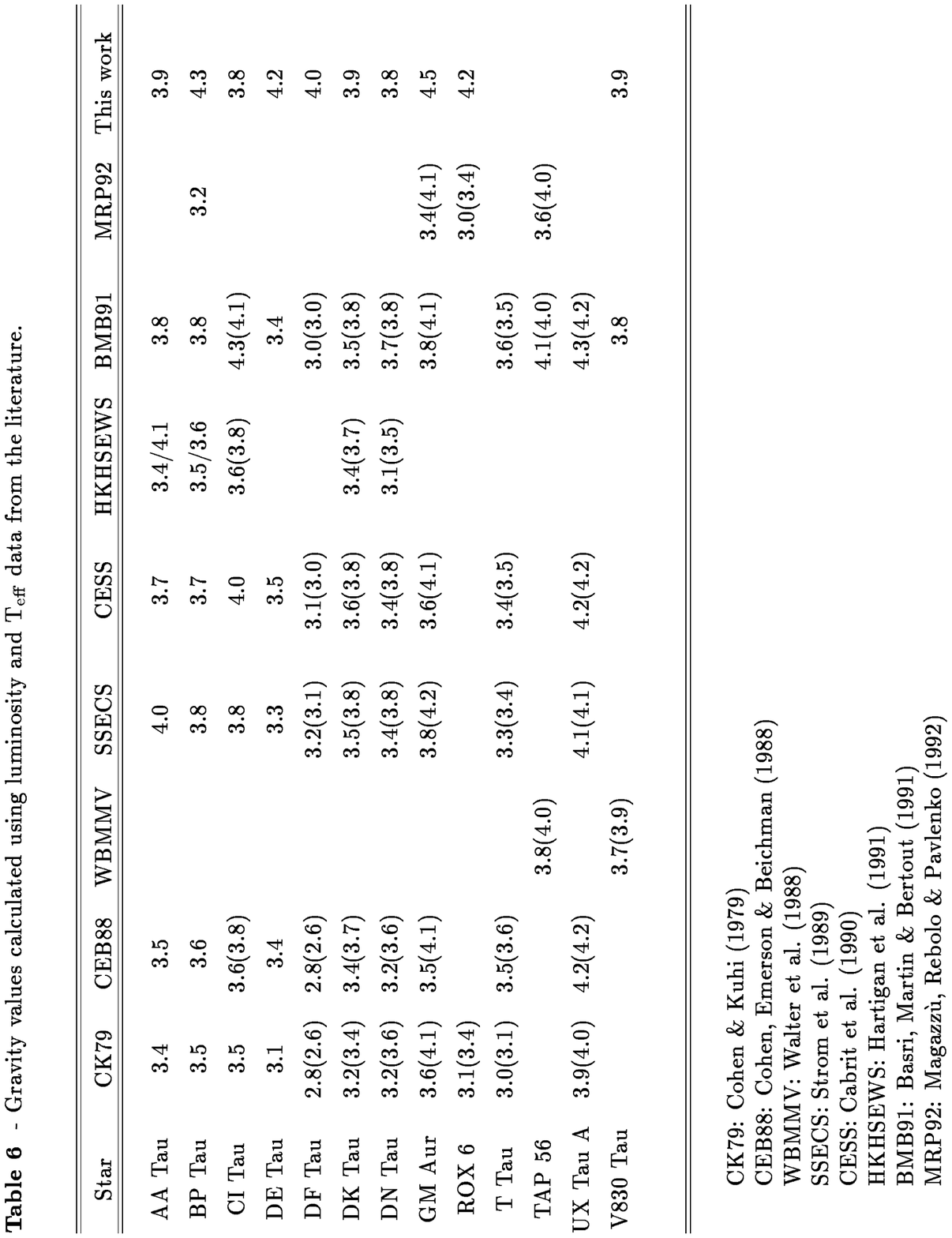}

\baselineskip 15 pt

\centerline{\bf Captions to the Figures}

\noindent $\bullet$ {\bf Figure 1.} The line pair used in the T$_{\rm
eff}$ determination, as seen in two Hyades stars of different spectral
types, showing their inverse temperature sensitivity.

\noindent $\bullet$ {\bf Figure 2.} The relation between line depth
ratio and (R-I)$_{Kron}$, using template stars data (Table 2)

\noindent $\bullet$ {\bf Figure 3.} Theoretical and observed relations
between line ratio and effective temperature. Long dashes stand for
log$\,g$ = 4.5, short dashes stand for log$\,g$ = 4.0, the dotted line
stands for log$\,g$ = 3.5 and the full line stands for log$\,g$ = 3.0.
The data are from the template stars listed in Table 2.

\noindent $\bullet$ {\bf Figure 4.} Synthetic spectra showing the response
of CaH lines to T$_{\rm eff}$ and log$\,g$.

\noindent $\bullet$ {\bf Figure 5.} Theoretical response of the line
depth ratios (VI 570.7/FeI 570.6 $nm$ and CaH 679.6/FeI 570.6 $nm$)
to T$_{\rm eff}$ and log$\,g$. The points show the final solutions
for the program stars. The correspondence is the following: triangle-DE
Tau, square-DF Tau, cross-AA Tau, asterisk-BP Tau, x-DK Tau,
losangle-CI Tau

\noindent $\bullet$ {\bf Figure 6.} Spectrum synthesis of rotational CaH
lines for DN Tau, GM Aur and ROX 6.

\vfill\eject

\topmongofig{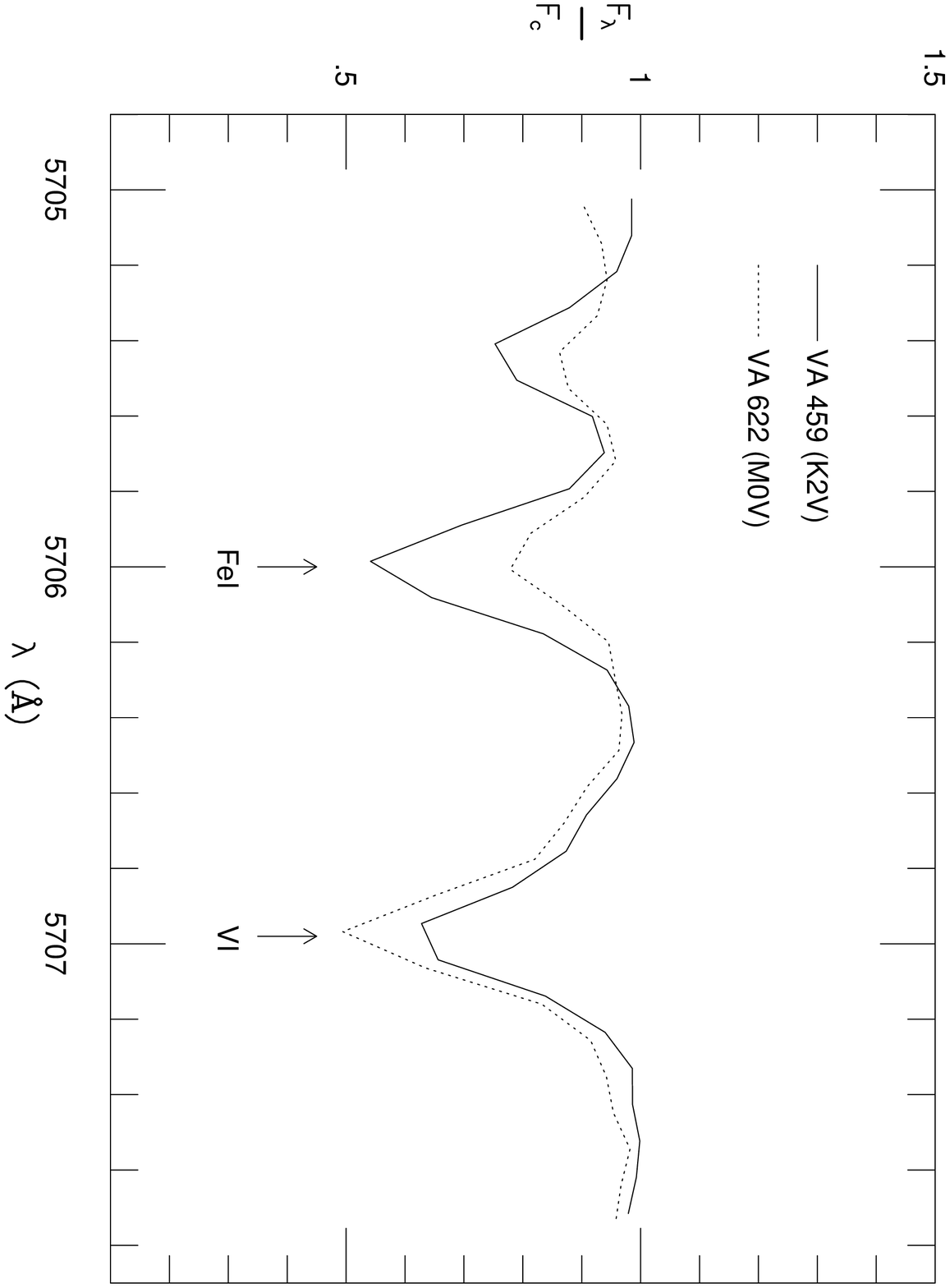}{1}

\topmongofig{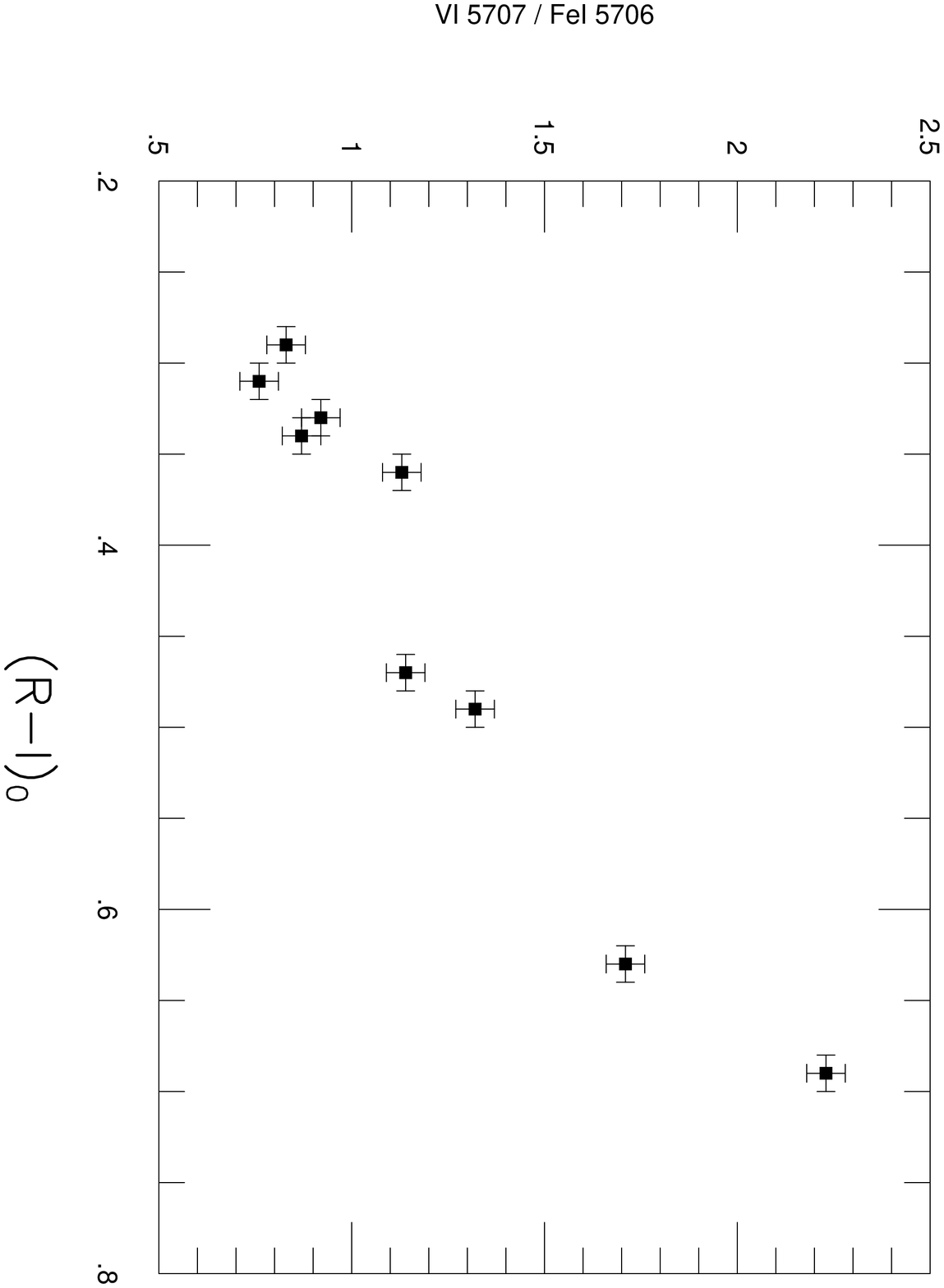}{2}

\topmongofig{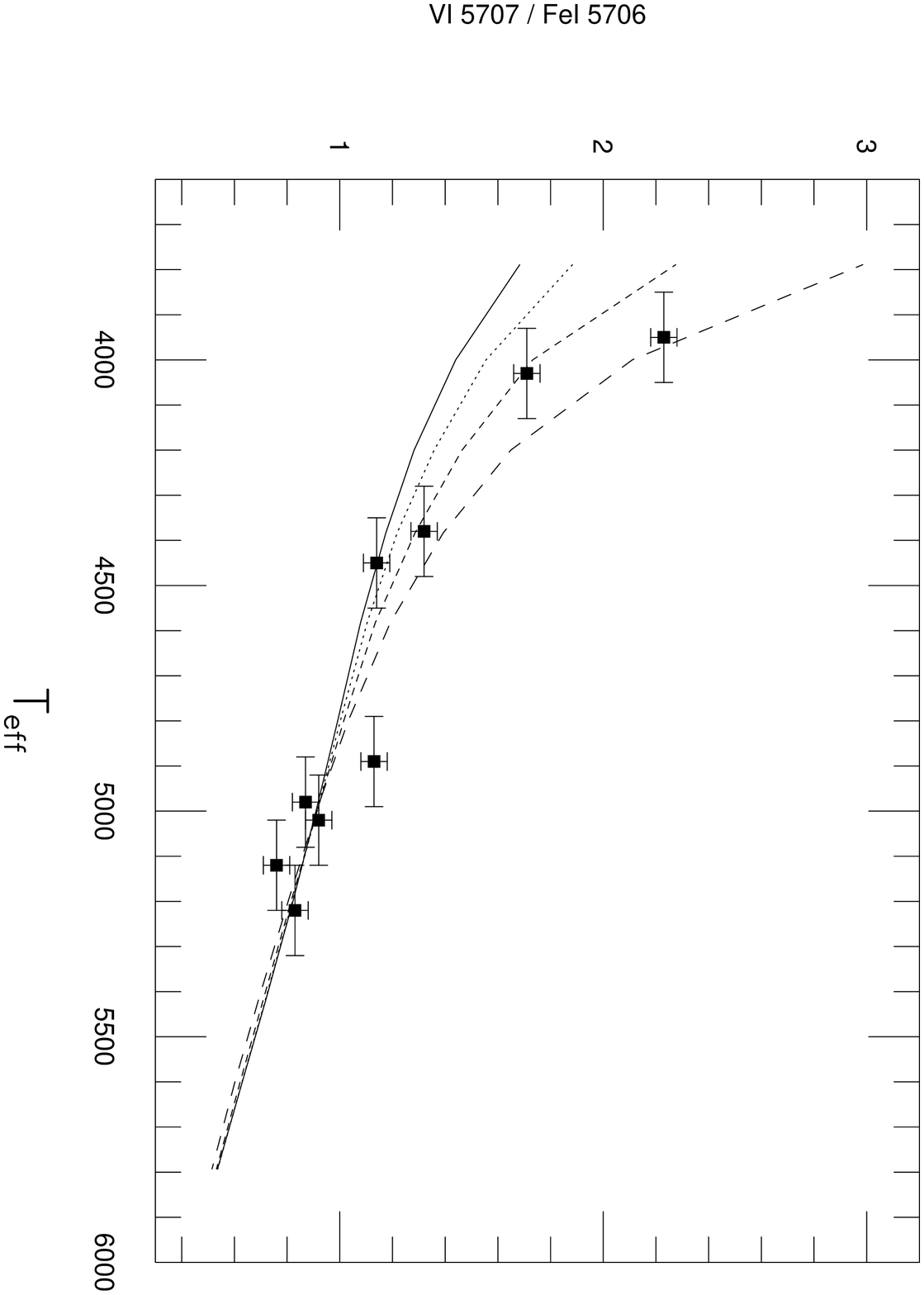}{3}

\topmongofig{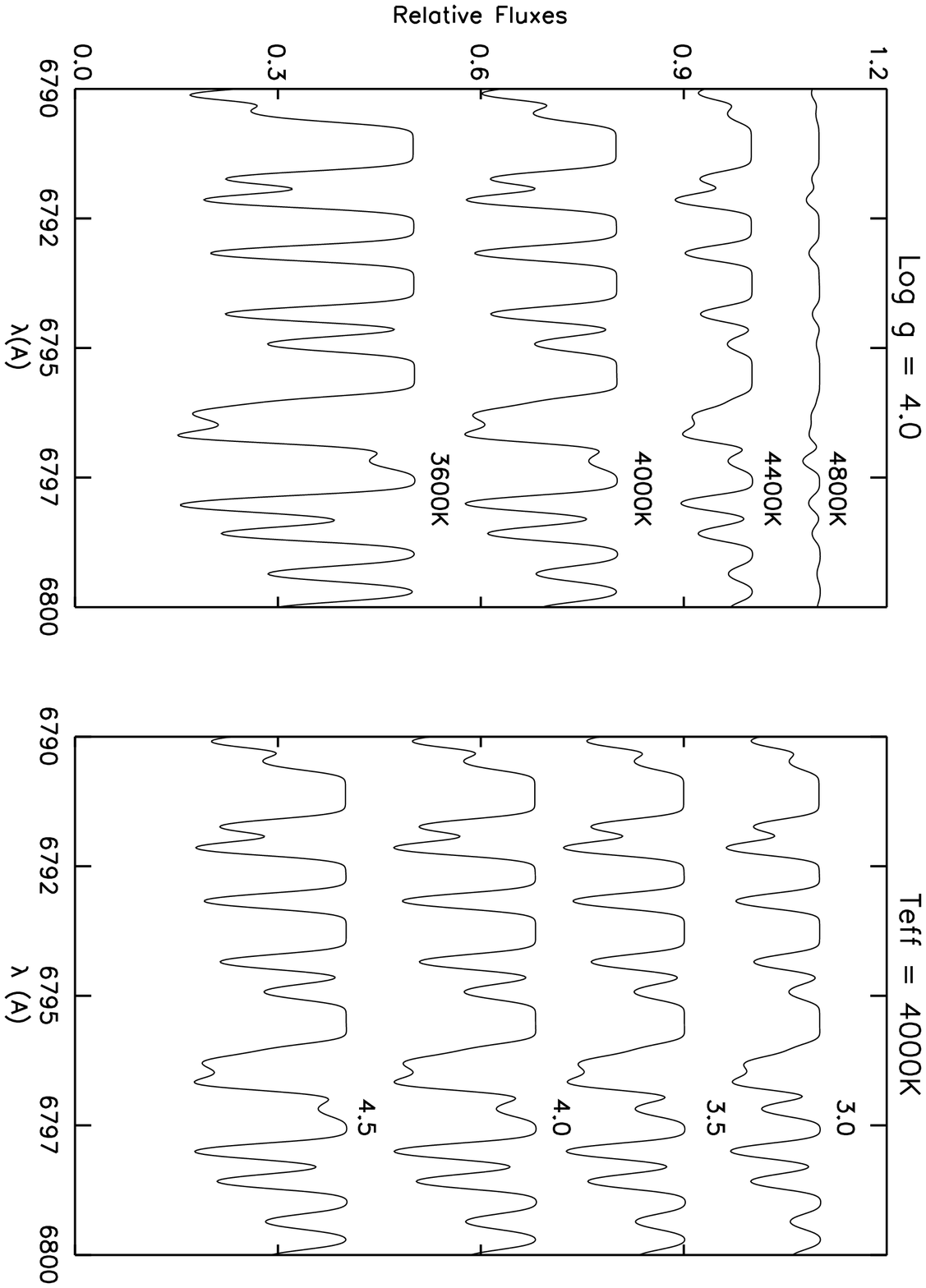}{4}

\topmongofig{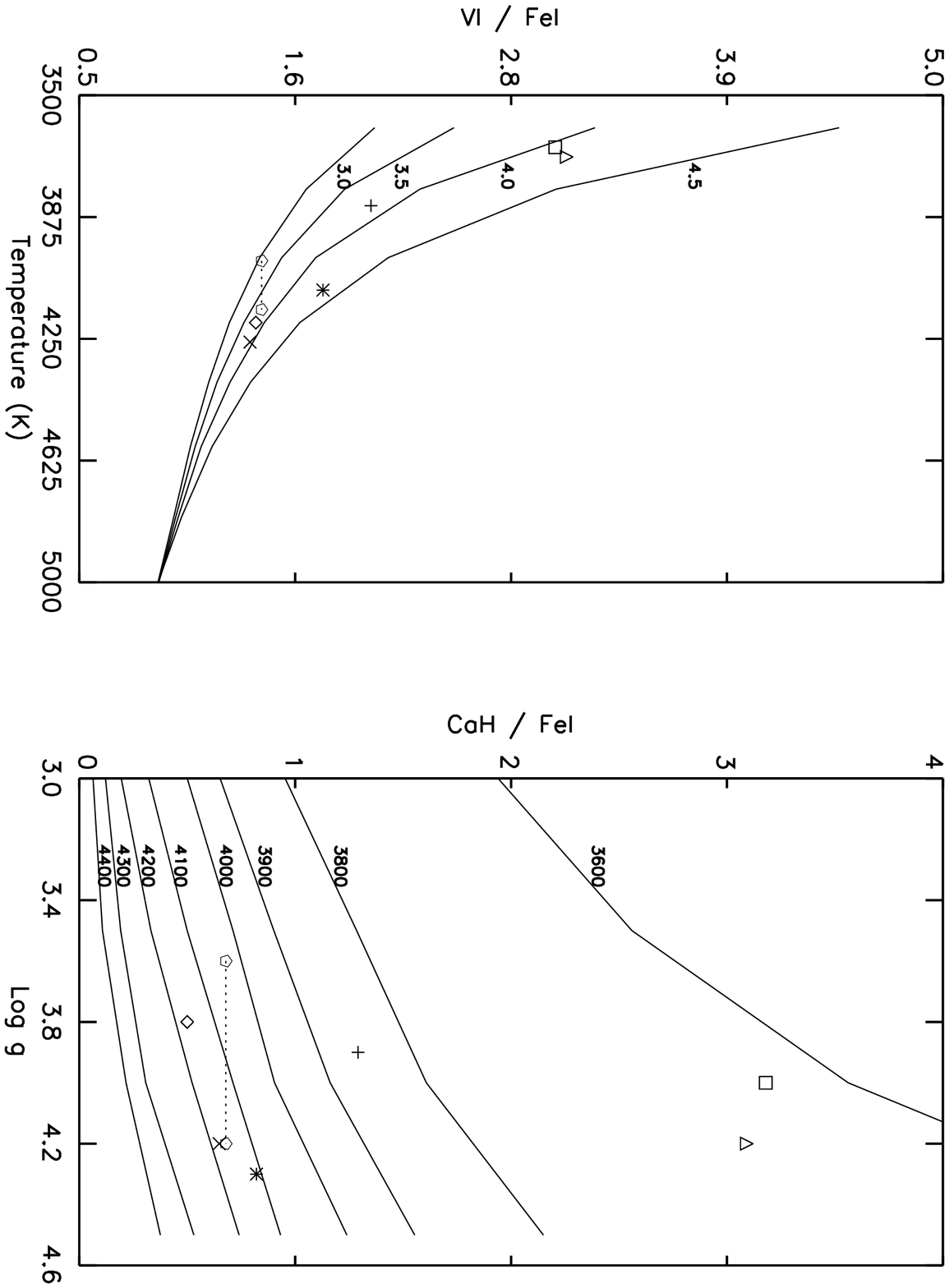}{5}

\pagemongofig{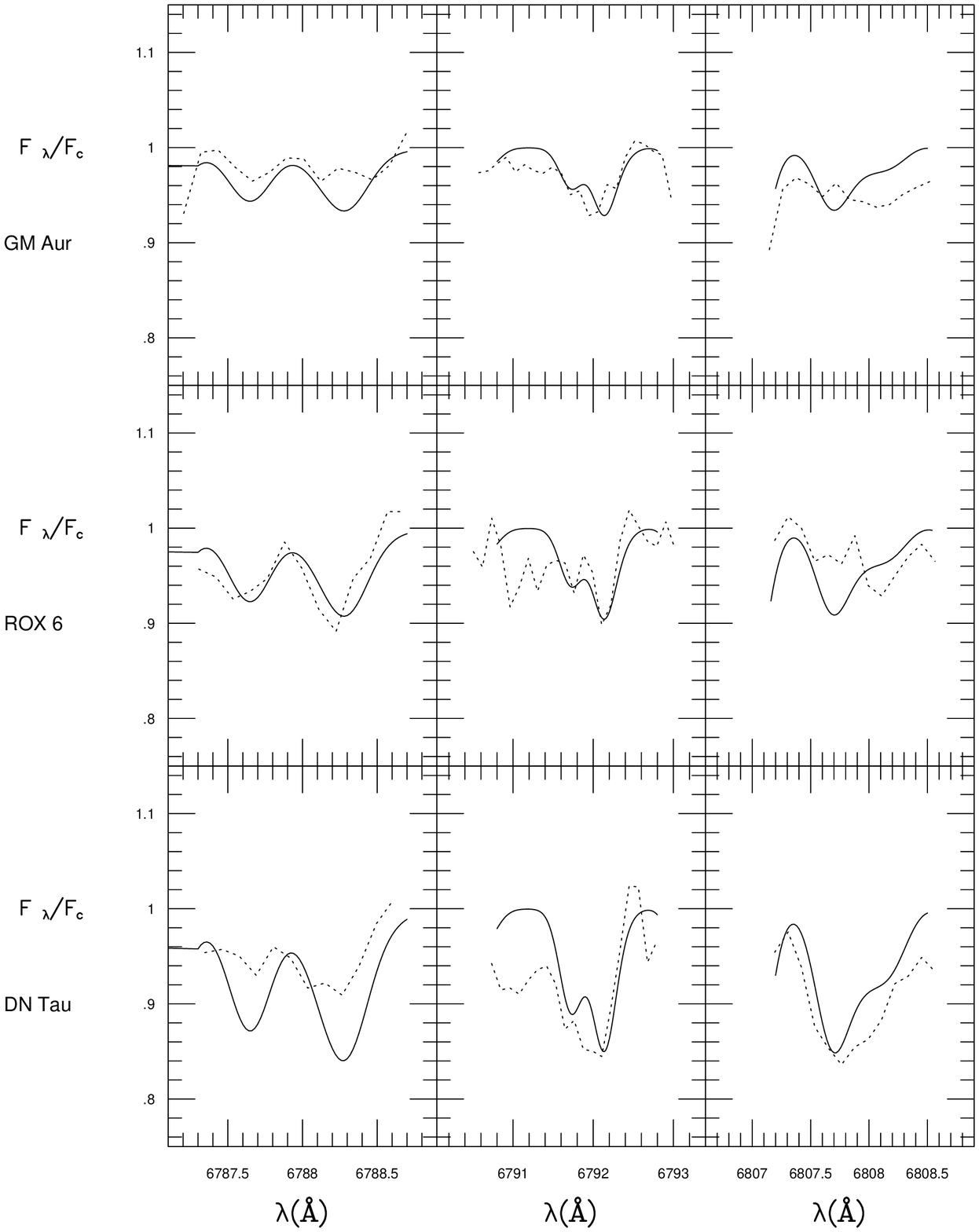}{6}

\pagemongofig{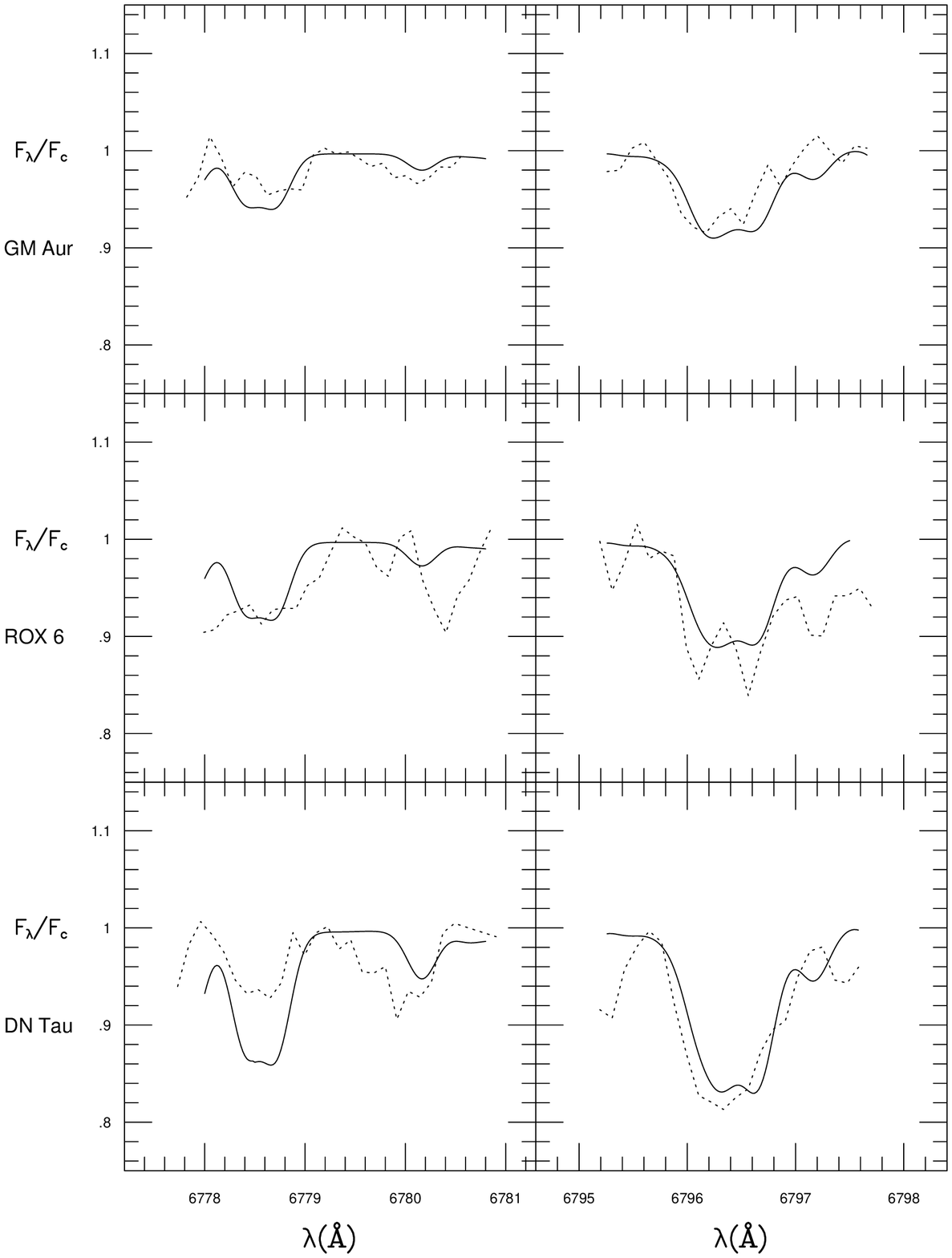}{6. Cont}

\end